\documentclass[authoryear,preprint,review,12pt]{elsarticle}

\usepackage{lineno}

\usepackage{graphicx}
\usepackage{amssymb} 
\usepackage{dcolumn}
\usepackage{lscape}

\newcommand{\arcsec}{$^{\prime\prime}$}

\newcommand{\cm}{\ensuremath{\rm\, cm^{-1}}}

\journal{Planetary and Space Science}

\begin{document}

\begin{frontmatter}



\title{Preparation for the Solar system observations with Herschel: Simulation of Jupiter observations with PACS}


\author[MPS]{Hideo Sagawa\corref{cor1}}
\cortext[cor1]{Corresponding author. Tel.: +49 5556 979217, fax.: +49 5556 979240.}
\ead{sagawa@linmpi.mpg.de}
\author[MPS]{Paul Hartogh}
\author[MPS]{Miriam Rengel}
\author[SRON]{Arno de~Lange}
\author[MPS]{Thibault Cavali\'{e}}

\address[MPS] {Max-Planck-Institut f\"{u}r Sonnensystemforschung, Max-Planck-str.\,2, 37191 Katlenburg Lindau, Germany} 
\address[SRON]{SRON Netherlands Institute for Space Research, Sorbonnelaan 2, 3584~CA Utrecht, The Netherlands}

\begin{abstract}
Observations of the water inventory as well as other chemically important species on Jupiter will be performed in the frame of the guaranteed time key project of the Herschel Space Observatory entitled {\it ``Water and related chemistry in the Solar system"}.
Among other onboard instruments, PACS (Photodetector Array Camera and Spectrometer) will provide new data of the spectral atlas in a wide region covering the far-infrared and submillimetre domains, with an improved spectral resolution and a higher sensitivity compared to previous observations carried out by Cassini/CIRS (Composite InfraRed Spectrometer) and by ISO (Infrared Space Observatory).

In order to optimise the observational plan and to prepare for the data analysis, we have simulated the expected spectra of PACS Jupiter observations. 
Our simulation shows that PACS will promisingly detect several H$_{2}$O emission lines.
As PACS is capable of spatially resolving the Jovian disk, we will be able to discern the external oxygen sources in the giant planets by exploring the horizontal distribution of water. 
In addition to H$_{2}$O lines, some absorption lines due to tropospheric CH$_{4}$, HD, PH$_{3}$ and NH$_{3}$ lines will be observed with PACS. Furthermore, owing to the high sensitivity of the instrument, the current upper limit on the abundance of hydrogen halides such as HCl will be also improved.  
\end{abstract}


\end{frontmatter}

\linenumbers


\section{Introduction} \label{intro}

One of the most outstanding results in previous infrared observations of Jupiter is the detection of water in its stratosphere, achieved by ISO (Infrared Space Observatory) \citep{feuchtgruber97,lellouch97}. 
ISO has also detected CO$_{2}$ in the Jovian atmosphere \citep{lellouch02} and these detections are regarded as a clear evidence for the existence of an external supply of oxygen into Jovian stratosphere. The observations of the Jovian water vapour with ISO have been followed by submillimetre (556.9\,GHz) heterodyne observations by SWAS (the Submillimeter Wave Astronomy Satellite) \citep{bergin00,lellouch02} and the Odin satellite \citep{cavalie08}.
In addition to monitoring the temporal variability since the ISO observations, these spectrally resolved measurements constrained the vertical distribution of the Jovian stratospheric water vapour.
Additional constraints on the possible external source have been obtained from the observations of CO, HCN and CS \citep[e.g.,][]{moreno03,griffith04}. 
In the current picture, most of the exogenous species originate from the Shoemaker-Levy\,9 (SL9) impacts in July 1994. 
However, other sources could be providers of oxygen, nitrogen and sulphur species as well in Jupiter as in the other giant planets. 
These potential sources are the following: i) the permanent interplanetary dust particle flux, ii) a local flux from the rings and icy moons via the magnetic field lines \citep{connerney86}, iii) sporadic large comet collision events. 
Discerning the relative contribution of each external source for each giant planet is a key issue, as it helps us in developing a full picture of various phenomena in the outer solar system such as the production of dust at large heliocentric distances, the role of the magnetosphere in the transport of exogenous material and the frequency of comet impacts.

The D/H ratio, which has been inferred from ISO measurements of the HD 37.7\,$\mu$m rotational line \citep{lellouch01}, is another key issue in Jovian atmospheric studies, because D/H on Jupiter and Saturn is considered as representative of the protosolar nebula while that on Uranus and Neptune represents the result of the mixing of the gases in the nebula with deuterium-rich icy cores \citep{feuchtgruber99}. 
Thus, the accurate comparison of D/H ratios on Jupiter and Saturn with Uranus and Neptune allows us to evaluate the D/H ratio of the protoplanetary ices embedded in the outer nebula, which is essential for a better understanding of the formation and evolution of the outer planets \citep{hersant01}. 

These unsettled issues will be entirely re-addressed by new and sensitive observations that will be carried out with the Herschel Space Observatory (launched on 14 May 2009) which covers a wide spectral range in the far-infrared and submillimetre domains (approximately 55--672\,$\mu$m). 
The guaranteed time key programme {\it ``Water and related chemistry in the solar system"} \citep{hartogh09}, also known as the HssO (Herschel Solar System Observations), is a dedicated project to observe water and other minor species in the solar system by using three onboard instruments: the Heterodyne Instrument for the Far Infrared \citep[HIFI;][]{degraauw08,HIFI2010}, the Spectral and Photometric Imaging REceiver \citep[SPIRE;][]{griffin08,SPIRE2010}, and the Photodetector Array Camera and Spectrometer \citep[PACS;][]{poglitsch08,PACS2010}. 
The focal plane units of these three instruments are kept cooled at the temperature range of 1.7--10\,K by using superfluid helium, and the bolometric detectors of SPIRE and PACS are further cooled down to 0.3\,K.\@
As a result of this sophisticated active cooling system and also of the instrumental development such as the state-of-the-art SIS (superconductor-insulator-superconductor) mixers on HIFI, Herschel enables observing with a sensitivity that has never been achieved before.
Among these instruments, PACS has the unique capability to obtain a spectral atlas in a wide wavelength range between 55 and 210\,$\mu$m with a resolving power of 940--5500.\@ 
This spectral resolution is rather low if compared to that of HIFI, and is not sufficient to resolve the molecular line shapes.
However, our previous preparatory study for the HssO Titan observations \citep{rengel09} has demonstrated that PACS spectroscopic observations can be employed to retrieve the vertical temperature profile and constrain the H$_{2}$O abundance in Titan's atmosphere by combining the observations of several lines of different opacity. 

In this study, we intend to extend our simulation study to the Jupiter observations. 
The Jupiter observations that are planned within the HssO project are described in Section~2.\@
The method used for the simulation is presented in Section~3, and expected results of PACS spectroscopy on Jupiter are discussed in Section~4.\@
The summary is given in Section~5.\@

\section{Planned Jupiter observations with Herschel} \label{obsplan}

Within the framework of the HssO project, PACS and HIFI are the proposed instruments for the Jupiter observations. 
SPIRE is not included in the proposal since its detector saturates when observing the bright continuum of Jupiter. 

PACS will be used in its spectroscopy mode for the HssO Jupiter observations. 
It is operated as an integral field spectrometer between 55 and 210\,$\mu$m over a field-of-view (FOV) of 47\arcsec $\times$47\arcsec (with a 5$\times$5 pixel array). 
The 47\arcsec\ FOV is comparable to the apparent disk diameter of Jupiter in its visible windows (36--45\arcsec), which enables us to obtain the disk resolved spectro-imaging of Jupiter with a single exposure. 
An image slicer is used to transform the signal observed with 5$\times$5 spatial pixels (1 pixel corresponds to 9.4\arcsec) at the focal plane into 1$\times$25 pixels which is the entrance slit for the grating spectrometer. 
The grating is operated in its first (102--210\,$\mu$m), second (72--98\,$\mu$m), or third (55--72\,$\mu$m) order. 
Two photoconductor arrays (16$\times$25 pixels, where 16 pixels are for the spectral dimension and 25 for the spatial) of red (102--210\,$\mu$m) and blue (55--98\,$\mu$m) channels are operated simultaneously, therefore it is possible to observe two grating orders together, either combination of 1$^{st}$\,+\,2$^{nd}$ or 1$^{st}$\,+\,3$^{rd}$. \@
The spectral resolving power varies in a range of 940--5500 depending on wavelength and grating order. 
The instantaneous spectral coverage (for 16 pixels) also varies from $\sim$0.15 to 1.05\,$\mu$m as well.

Two observing modes, {\it line scan} and {\it range scan} modes, are prepared for the PACS spectroscopy. 
The line scan mode is employed when one observes one or several unresolved narrow line features in a fixed wavelength range of around 1\,$\mu$m, while the range scan mode is used to observe any wider wavelength range.
High and low spectral sampling configurations are available for these line and range scans. 
Molecular lines will be sampled with at least two spectral pixels even at the low sampling rate. 
PACS can obtain a full range spectrum between 55 and 210\,$\mu$m (so-called {\it SED} mode) in one hour by combining two low sampling range scans ($\sim$1300 seconds for observing the 55--73 and 102--146\,$\mu$m ranges, and $\sim$2400 seconds for 70--105 and 140--210\,$\mu$m). 
The expected noise equivalent spectral radiance (NESR) for a single repetition of the SED range scan observation is 0.8--4.0\,Jy varying as a function of the wavelength (the best sensitivity is achieved at the wavelength region of 110--140\,$\mu$m). It is to be noted that this sensitivity estimation is valid when observing a low/moderate brightness source, and it is subject to change for the Jupiter observations which contain a very bright continuum emission.
The optimization of the instrumental setup of PACS for the bright target is now under way by the PACS Instrument Control Centre.
More details of the instrumental information can be found in the PACS observer's manual\footnote{http://herschel.esac.esa.int/Docs/PACS/pdf/pacs\_om.pdf}.

The proposed Jupiter observations with PACS are: 
\begin{itemize}
\item The high signal-to-noise ratio (SNR) {\it SED range scan} from 55 to 210\,$\mu$m, which aims at not only the observation of numerous water lines but also the first detection of hitherto unobserved molecules. 
An SNR of at least 100 with respect to the continuum is required in the HssO proposal.
\item A dedicated {\it line scan} spectroscopy observation of the 66.4\,$\mu$m (4512\,GHz) H$_{2}$O line with the high spectral sampling mode.
\end{itemize}
In order to search for temporal variability, these observations (as well as the undermentioned HIFI observations except the mapping one) will be repeated three times during the lifetime of Herschel ($\sim$3.5 years).

As the explicit instrumental sensitivity of PACS for the very bright source is not confirmed yet, we do not have the practical number of the expected SNR for the abovementioned observation programmes. 
According to the recent results of the PACS Performance Verification phase, an SNR of at least 1000 with respect to the continuum is currently expected for the Jupiter SED range scan observation (personal communications with Helmut~Feuchtgruber).\@ Instead, in this study we investigate the possible outcome of the PACS Jupiter observation for the SNR required in the guaranteed time proposal i.e. SNR $>$100 with respect to the continuum.


Although the focus of the present study is the PACS observations, we briefly describe here about the Jupiter observations with HIFI in the HssO project as they provide essential information of the vertical profile of the water vapour. 
HIFI is composed of two HEB (hot electron bolometer) and five SIS mixer bands that cover the wavelength range of 157--212 and 240--625\,$\mu$m (corresponding to 1910--1410 and 1250--480\,GHz, respectively). 
As backends, two spectrometers are available; the High Resolution Spectrometer (HRS, an auto-correlator spectrometer) and the Wide Band Spectrometer (WBS, an acousto-optical spectrometer).\@
HRS enables the instantaneous coverage of 250\,MHz with spectral resolutions ranging from 0.14 to 0.54\,MHz, or coverage of 500\,MHz with 1.1\,MHz resolution, while WBS covers 4\,GHz with a spectral resolution of 1.1\,MHz. 
The beam size of HIFI is 11--44\arcsec\ depending on the observing wavelength. 

With HIFI, the following Jupiter observations are proposed: 
\begin{itemize}
\item High SNR observations of the strong H$_{2}$O lines at 557, 1097, and/or 1670\,GHz (538.2, 273.2 and/or 179.5\,$\mu$m).\@ From their line shapes, the vertical profile of H$_{2}$O will be retrieved.
\item Rough mapping (10 points over the Jovian disk) of the 1670 or 1717\,GHz H$_{2}$O line (the beam size of HIFI is $\sim$12\arcsec\ at these frequencies). Combined to the PACS observations, these observations will be used to determine the spatial distribution of H$_{2}$O in the atmosphere of Jupiter. 
\item The 1882\,GHz (159.3\,$\mu$m) CH$_{4}$ line observation.\@ CH$_{4}$ in Jupiter is considered to be uniformly mixed in altitude and its abundance is well known. Therefore, its spectral line shape will provide us with the thermal profile.  
\end{itemize}


\section{Simulation of the PACS observations} \label{simulation}

Our simulation package is based on the general forward and inversion model called MOLIERE-v5 \citep{urban04}. 
The forward model consists of a radiative transfer model for the Jovian atmosphere and an instrumental part where the atmospheric spectrum is convolved with Herschel's main beam pattern and PACS spectral response. 
We use a line-by-line radiative transfer model with a multi-layered spherical atmosphere which the pressure levels ranging from 6\,bar to 2$\times$10$^{-4}$\,mbar. 
The reference temperature profile is based on the radio occultation experiments of Voyager\,2 \citep[Table~1 in][]{lindal92} and the {\it in situ} measurement of the Galileo entry probe \citep{seiff98}.
For simplicity, we use a smoothed temperature profile for the stratosphere (Fig.~1(a)). 
NH$_{3}$, PH$_{3}$, CH$_{4}$, CO, HCN, H$_{2}$O, HCl and HD are considered in the line opacity calculations. 
Adopted vertical profiles of their mixing ratios are shown in Fig.~1(b).\@ 
The profiles of NH$_{3}$ and PH$_{3}$ are based on the reference model of \citet{nixon07}. 
The CH$_{4}$ profile is based on the photochemical model of \citet{moses05}.
Profiles of H$_{2}$O, CO and HCN are derived from the observations of the SL9 impact \citep{lellouch97, bezard02, moreno03}.
For HCl, an upper limit of 2.3\,ppb has been retrieved from the Cassini/CIRS (Composite InfraRed Spectrometer) measurements \citep{fouchet04}. In this study, we use three HCl profiles having vertically constant mixing ratios of 2.3, 1.15 and 0.23\,ppb, respectively. 
The spectroscopic parameters are derived from the HITRAN 2008 compilation \citep{HITRAN08} except for PH$_{3}$ and HD.\@ 
Parameters for PH$_{3}$ are derived from the GEISA 2003 compilation \citep{GEISA03}.
For HD, the parameters from the CDMS compilation \citep{CDMS2005} are used.
We have used two line shape functions depending on the pressure: The Voigt line shape is used at the pressures where the pressure broadened half width is less than 40 times of the Doppler half width. At higher pressures, the Van Vleck-Weisskopf line shape is adopted alternatively. 
The pressure broadening line widths are derived from those listed in the HITRAN and GEISA databases which are appropriate for the Earth atmosphere, because there are few laboratory and theoretical works on those induced by Hydrogen and Helium in particular at the submillimetre/far-infrared spectral region.
The collision induced absorption coefficients of H$_{2}$ and He mixtures are included by using the formulation of \citet{borysow85, borysow88}. 
Possible additional opacity due to clouds and hazes was not considered in this preparatory work.
The distance between Herschel and Jupiter was set to 5\,AU, corresponding to an apparent disk diameter of 39\arcsec. 
We assume that the telescope is pointing to the centre of the disk, and calculate the spectrum detected by the central pixel of the PACS detector array. 
The radiative transfer calculation was performed with a very fine spectral grid, containing all the transition frequencies and the nearby line wing frequencies of the species mentioned previously, in order to evaluate the line shape of the molecular emission correctly.
The synthesised spectrum after the radiative transfer calculation is convolved by the PACS instrumental function which accounts for its finite spectral resolution ($\lambda / \Delta\lambda$\,=\,940--5500). \@
As described in the previous section, the spectral samplings (i.e. frequencies at the final spectral grid to be synthesised) of PACS vary depends on the observing mode. In order to examine the best possible performance of PACS full range scan observations, we used an oversampled spectral gridwhich includes all the molecular transition frequencies.

In this study, we simulate the PACS Jupiter spectrum with an SNR of 100 with respect to the continuum emission at the reference wavelength of 200\,$\mu$m. For other wavelengths, we scaled the corresponding NESR from the standard SED range scan observation in accordance with its frequency dependency. This assumed NESR results in the PACS Jupiter spectrum having an SNR as high as $\sim$1200 with respect to the continuum at the wavelength around 110\,$\mu$m, and lower than 100 at the end wavelengths of each grating order. 

The sensitivity of PACS measurements to the atmospheric parameter of interest (such as the vertical distribution of H$_{2}$O) is examined by calculating the weighting functions. 
The weighting function is defined as the derivative of the forward model with respect to the atmospheric parameter at each measurement frequency channel.

\section{Expected outcome of PACS observations} \label{results}

\subsection{H$_{2}$O measurement with PACS}

As previously described in Section~2, the primary objective of the HssO project with respect to Jupiter observations is to determine the vertical and horizontal distribution of water in the Jovian stratosphere. 
We therefore start with examining the sensitivity of PACS full range scan spectroscopy observations to the water vapour abundance.

Figures~2--3 show the synthetic full range spectra of Jupiter considering PACS spectral resolution.
For comparison, the synthetic spectrum before convolution with the instrumental function is also shown there. 
The continuum emission comes from the collision induced opacity of the H$_{2}$ and He atmosphere, and the broad absorption line features of tropospheric NH$_{3}$. 
Although a large number of the roto-vibrational transitions of the considered species are present in the frequency region of interest (see ticks on the middle panels of Fig.~2--3), most of them are severely broadened due to PACS finite spectral resolution and become indistinctive from the continuum. 
However, by checking the line-to-continuum ratio, a couple of molecular lines including H$_{2}$O show their emission/absorption amplitude being as large as $\sim$7\,\% of the continuum. 
Under the assumed NESR conditions (i.e. SNR of $\sim$100--1200 with respect to the continuum), those lines will certainly be detected from the set of PACS observations.

While ISO/LWS (Long Wavelength Spectrometer) observations have resulted in the detection of the two strongest H$_{2}$O emission lines at 66.4 and 99.5\,$\mu$m in the PACS spectral range, Cassini/CIRS has not succeeded in detecting any H$_{2}$O line in the Jovian atmosphere due to the limitations in terms of spectral resolution and sensitivity.
In case of the assumed PACS measurements with an SNR of 100 at 200\,$\mu$m, our calculations suggest that many more H$_{2}$O lines, up to 22 in the presented case, become detectable at the 5-$\sigma$ level for the first time in this far-infrared/submillimetre domain (indicated by arrows on the lowermost panels of Fig.~2--3).\@

Figure~4(a),(b) shows the weighting functions, with respect to the assumed H$_{2}$O vertical profile, for the 66.4\,$\mu$m H$_{2}$O line before and after considering PACS spectral resolution, respectively. 
Although the 66.4\,$\mu$m line when computed at infinite spectral resolution show sensitivity to a wide altitude range, all the weighting functions show their peaks at the same pressure level of 10--20\,mbar after smoothing to PACS spectral resolution. 
This is also the case for all other H$_{2}$O transitions in the PACS spectrum (Fig.~4(c)).
This means that we will be able to measure the H$_{2}$O abundance in this single altitude layer from the PACS multiple line measurements. 
From the perspective of exploring the vertical profile of H$_{2}$O in the Jovian stratosphere, the observations planned with HIFI are required.

The most interesting Jovian water observation with PACS is the investigation of a possible variation of its abundance as a function of latitude. 
Recent Cassini/CIRS observations have revealed that HCN and CO$_{2}$, which were both injected by the SL9 impact, are distributed over the planet in quite different ways \citep{lellouch06}.
Thus, it is of interest to investigate the latitudinal variations in the H$_{2}$O field as probably most of the stratospheric water originates also from this impact \citep{lellouch02,cavalie08}. 
This observation is possible because of the 5$\times$5 pixel coverage of PACS, and will be the first ever mapping of Jupiter at H$_{2}$O frequencies. 

\subsection{Detectability of minor species}

In addition to H$_{2}$O lines, several rotational lines of CH$_{4}$ are expected to be detected by PACS.\@
In Jupiter, CH$_{4}$ is distributed almost uniformly in both the troposphere and stratosphere.
Unlike HIFI, PACS is not sensitive to the narrow emission line of CH$_{4}$ coming from the stratosphere but will be able to observe accurately the absorption caused by tropospheric CH$_{4}$ (Fig.~5(a)). 

For the currently undetected species, the detection or stringent constraints on their abundances are expected with the PACS observations, since PACS achieves a higher spectral resolution and sensitivity than Cassini/CIRS. As an example, we present the detectability of HCl with PACS in this study.
We have calculated the spectra for different mixing ratios; 2.3\,ppb (current upper limit, and the spectra are shown in the Fig.~2--3), 1.15\,ppb and 0.23\,ppb. 
From the lowermost panels of Fig.~2--3, it is expected that the HCl\,(4--3) transition at 119.9\,$\mu$m is observed with the highest SNR among other HCl transitions, and can be used for acquiring the lower upper limit on the HCl abundance. 
Even in the case of the lowest assumed abundance, the HCl\,(4--3) spectral feature is likely observed at the 5-$\sigma$ level (Fig.~5(b)).\@ 
HCl and other hydrogen halides are theoretically expected to condense into solid ammonium halide salts (e.g., NH$_{4}$Cl) around the cold Jovian upper troposphere; therefore, the hydrogen halide vapours may only be detectable if they are present at disequilibrium for any reason. 
PACS observations of the HCl and other hydrogen halide abundances will improve the thermochemical modeling including the NH$_{3}$ cloud microphysics of Jupiter.  

Fig.~5(c),(d) shows the synthetic spectra of HD rotational transitions R(0) and R(1) located at 112.1 and 56.2\,$\mu$m (2675 and 5331\,GHz), respectively, assuming HD/H$_{2}$ ratio of 4.8$\times$10$^{-5}$ \citep{lellouch01}. 
Although the R(1) transition has a $\sim$2.5 times higher line opacity than R(0), the R(0) transition might be the best target for PACS to improve the accuracy of the previously measured HD/H$_{2}$ ratios in the giant planets from the view point of the sensitivity of the PACS spectrometer. In fact, the HD R(0) transition lies at the most sensitive wavelength region of the PACS spectrometer, while the position of the R(1) line is very close to the end of PACS spectral range where its sensitivity substantially degrades. 
Moreover, the HD R(1) line in Jupiter is located in intense NH$_{3}$ manifolds which makes difficult to determine the continuum level accurately (here continuum is referenced to the H$_{2}$ and He collision induced absorptions and to the NH$_{3}$ opacity). 
Under the assumed NESR conditions, the 10-$\sigma$ detection of the HD R(0) line is expected even though absorption depth is estimated as small as 1\,\% of the continuum.
  

\subsection{Sensitivity to tropospheric PH$_{3}$ and NH$_{3}$}

PACS will be sensitive to the PH$_{3}$ abundance in the upper troposphere, from $\sim$600\,mbar to 80\,mbar (Fig.~6(a)). 
PH$_{3}$, particularly in the upper troposphere, is considered to be an important probe for investigating the vertical and meridional atmospheric transport in the giant planets. 
PH$_{3}$ is enriched in the upper troposphere with respect to predictions by thermodynamic equilibrium models.
This has been interpreted as an evidence for rapid upward transportation of air from the deep interior as explained below \citep[e.g.,][]{barshay78}.
PH$_{3}$ is convectively transported from the deep hot atmosphere into the upper troposphere where it is in disequilibrium with the cold ambient temperatures.
Because the equilibrium reaction (4PH$_{3} +$6H$_{2}$O\,$\rightarrow$\,P$_{4}$O$_{6} +$12H$_{2}$) is very slow compared to the diffusion time scale, PH$_{3}$ actually remains in disequilibrium leading to the observed enhanced abundances.

Recent studies with Cassini/CIRS and ground-based telescopes observations \citep{irwin04,fletcher09b,fletcher09a} have revealed an enhancement of the PH$_{3}$ abundance at the equator compared to the neighbouring equatorial belts and mid-latitudes. 
Unfortunately, PACS spatial resolution is too low compared to the resolution of these previous observations so that PACS may not be able to observe such a fine latitudinal inhomogeneity; but still, PACS observations will be important in that they will extend the current results with a new observing wavelength region (currently, PH$_{3}$ is investigated using the CIRS spectra at mid-infrared 1200\,\cm\ region).
This will confirm previous findings and/or lead to new insights.

Tropospheric NH$_{3}$ will also be measured and, by analysing its very broad absorption lines, the bulk abundance of NH$_{3}$ at the level from $\sim$800\,mbar to 100\,mbar will be determined (Fig.~6(b)). 
This will improve our knowledge of the enrichment of nitrogen in Jupiter with respect to the solar value (Jovian nitrogen is enriched by a factor of $\sim$3 with respect to the solar value, \citep{wong04}) along with carbon and sulphur, which have been first confirmed by the Galileo probe measurements \citep{owen99}. 

\section{Summary} \label{summary}

The Herschel Space Observatory will bring new insights into the origin of stratospheric water as well as other controversial issues in the Jovian atmosphere, in the frame of the guaranteed time key programme entitled {\it ``Water and related chemistry in the Solar system"}. 

To optimise the observational plan and to prepare for the data analysis, we have simulated the expected Jupiter spectra as observed with PACS in the full range scan mode by developing a forward model that takes into account the radiative transfer in the Jovian atmosphere and PACS instrumental characteristics. 
Several H$_{2}$O emission lines will be detected for the first time in the wide far-infrared/submillimetre domains, and some absorption lines due to the tropospheric CH$_{4}$, PH$_{3}$ and NH$_{3}$. 
As PACS is capable of spatially resolving Jupiter, we will be able to newly constrain the origin of water in Jupiter by determining its horizontal distribution. 
Furthermore, the constraints on the abundances of hydrogen halides such as HCl will be also improved with respect to the current upper limit derived from the Cassini/CIRS measurements. 
For the observation of HD with PACS, our simulations suggest its rotational transition R(0) as the possible target to observe by making effective use of the most sensitive wavelength of the PACS spectrometer.  



\bibliographystyle{elsarticle-harv}
\bibliography{PSS_2869_accepted_Sagawa}

\clearpage

\begin{figure}
\begin{center}
\includegraphics[height=1.95in]{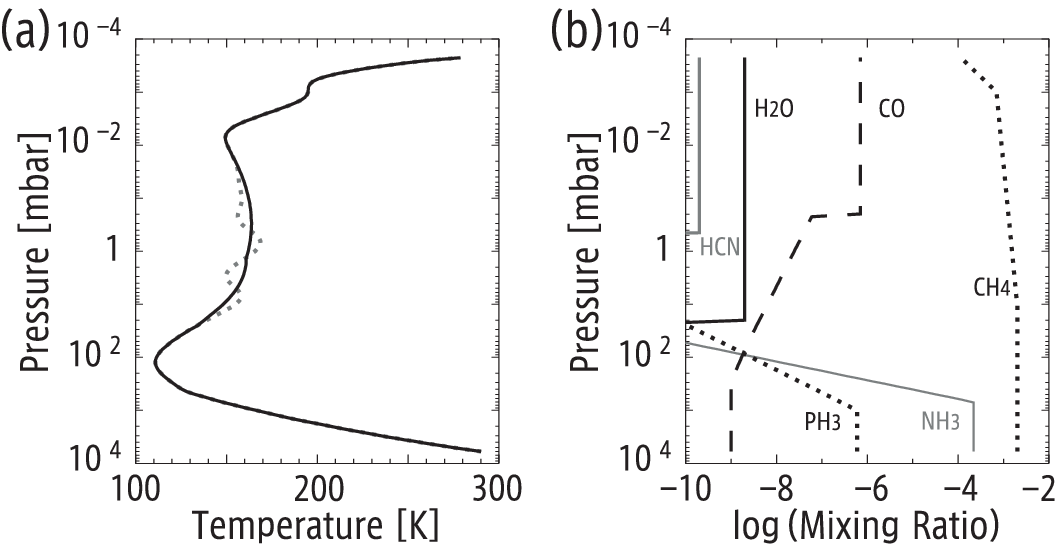}\\
\caption{{\textbf (a)} Temperature and {\textbf (b)} molecular mixing ratio profiles used in this study.
The dotted temperature profile represents the result obtained by the Galileo entry probe \citep{seiff98}.
The mole fractions of H$_{2}$ and He are set to 0.863 and 0.134, respectively, as measured by the Galileo probe \citep{niemann98} and not shown in the plot.}
\label{TP_VMR}
\end{center}
\end{figure}


\clearpage

\begin{figure}
\begin{center}
\includegraphics{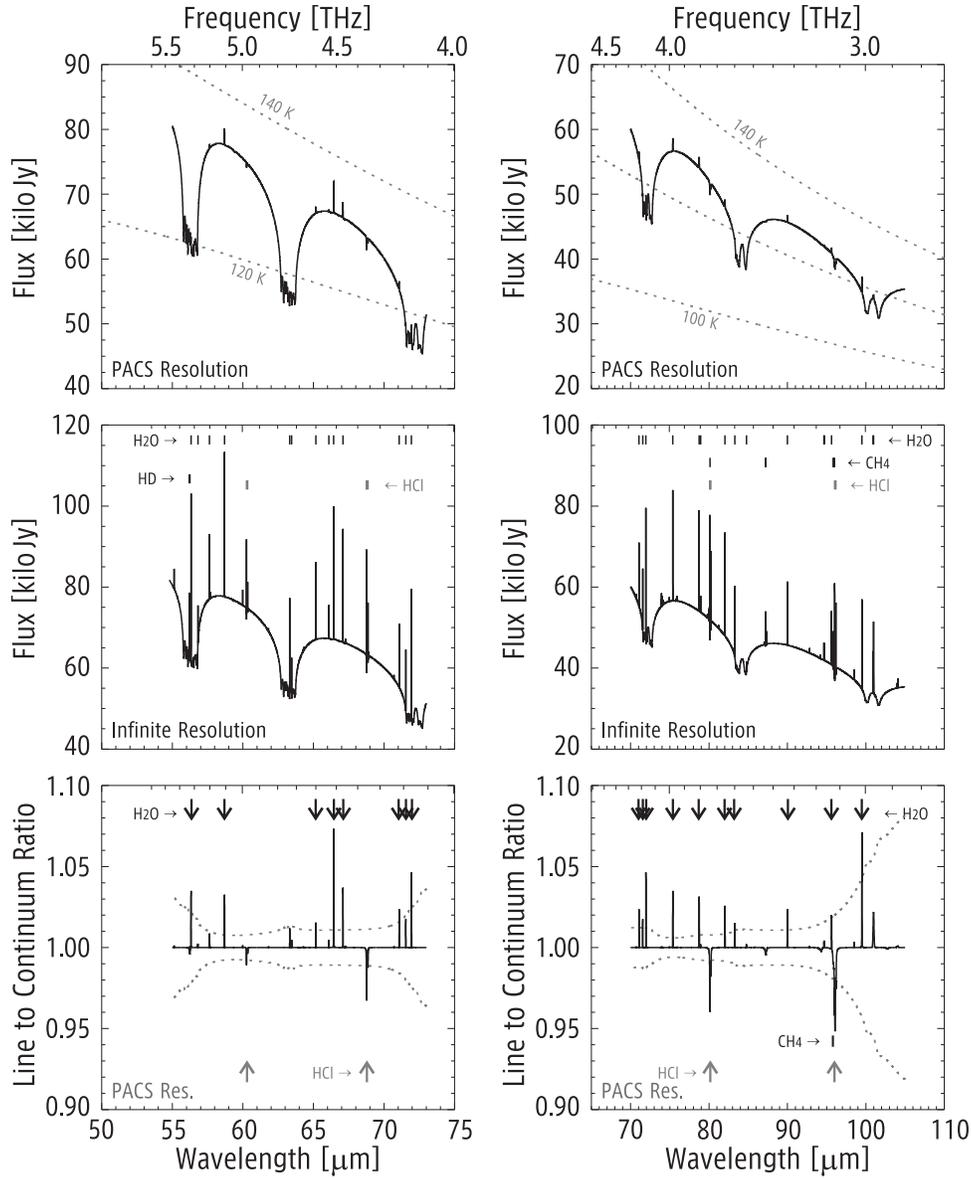}\\
\caption{Synthesised full range spectrum of PACS Jupiter observation for the blue channel. 
Left and right columns correspond to 3$^{rd}$ and 2$^{nd}$ order of the grating, respectively. 
{\textbf (a)}: The synthetic spectrum considering PACS spectral resolution. 
The amplitude is shown in terms of flux density for one spatial pixel. 
The dotted curves represent the level of flux emitted by a black body of a constant temperature (100, 120 and 140\,K). 
{\textbf (b)}: Infinite spectral resolution spectrum before convolution with PACS instrumental function. Small ticks show the line positions for molecules of interest.
{\textbf (c)}: The line-to-continuum ratio spectrum where the continuum emission is referenced to the H$_{2}$ and He collision induced absorptions and to the NH$_{3}$ opacity. 
The dotted curves represent the 5-$\sigma$ level calculated from the assumed NESR.
}
\label{PACS_full_blue}
\end{center}
\end{figure}

\clearpage

\begin{figure}
\begin{center}
\includegraphics{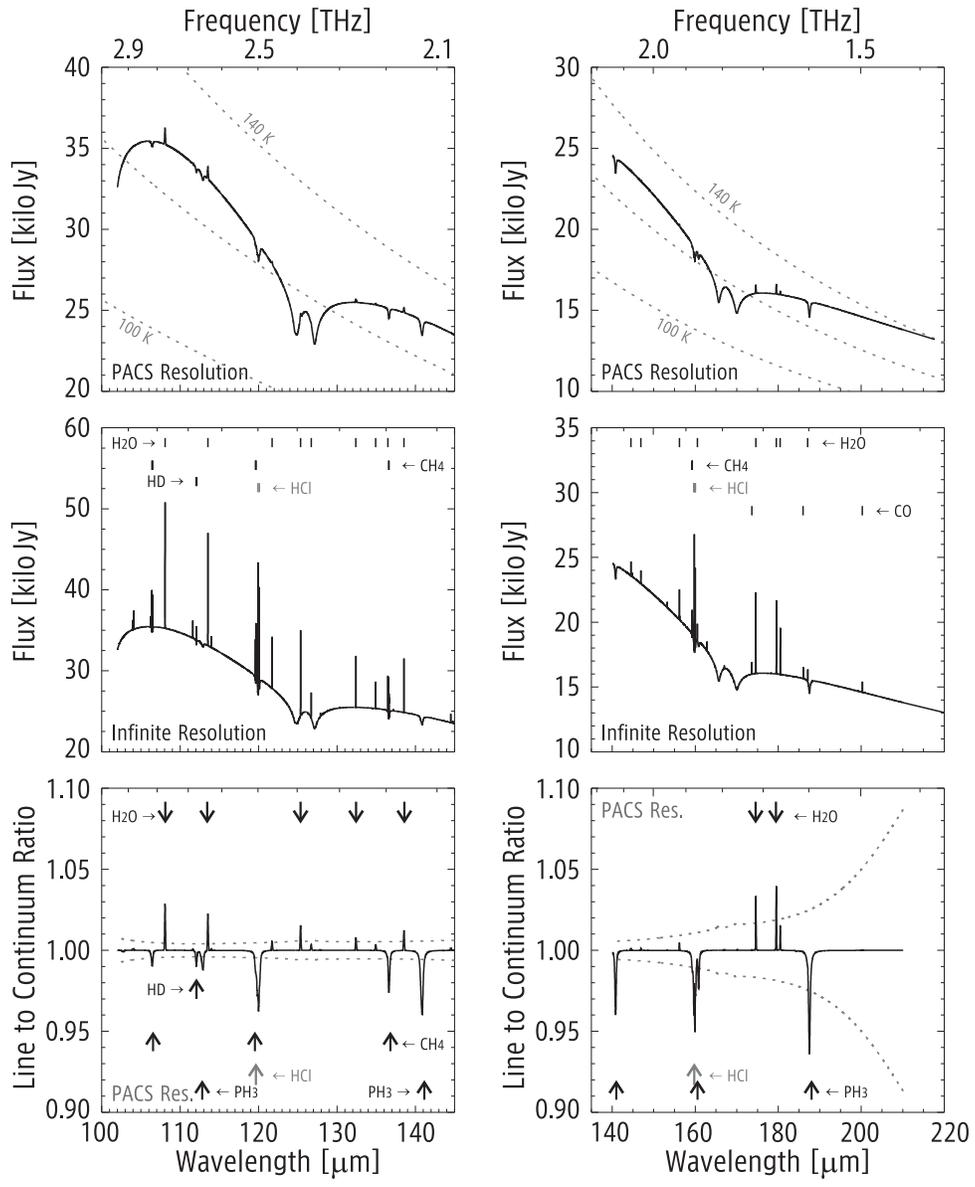}\\
\caption{Same with Fig.~2, but for the red channel (1$^{st}$ order of the grating).}
\label{PACS_full_red}
\end{center}
\end{figure}

\clearpage

\begin{figure}
\begin{center}
\includegraphics{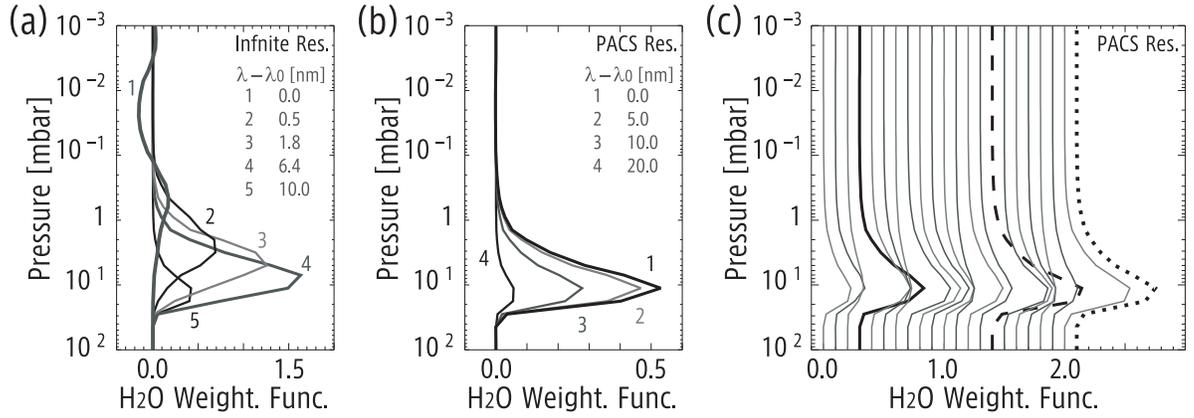}\\
\caption{Weighting functions with respect to the H$_{2}$O abundance, for the 66.4\,$\mu$m line ((a) \& (b)) and for different H$_{2}$O transitions indicated by the arrows at the lowermost panels of Fig.~2--3.\@ 
The panel (a) refers to the case without convolution with the PACS instrumental function, whereas the panels (b) \& (c) correspond to the case with this convolution taken into account.
In the panels (a) \& (b), each curve represents the weighting function at different wavelength offset from the line centre. 
In the panel (c), each weighting function is plotted with a certain offset value, and the curves for the three highest SNR lines at $\lambda$\,=\,66.4, 99.5 and 179.5\,$\mu$m are highlighted with bold, dashed and dotted lines, respectively, for clarity.
}
\label{H2O_WF}
\end{center}
\end{figure}

\clearpage

\begin{figure}
\begin{center}
\includegraphics{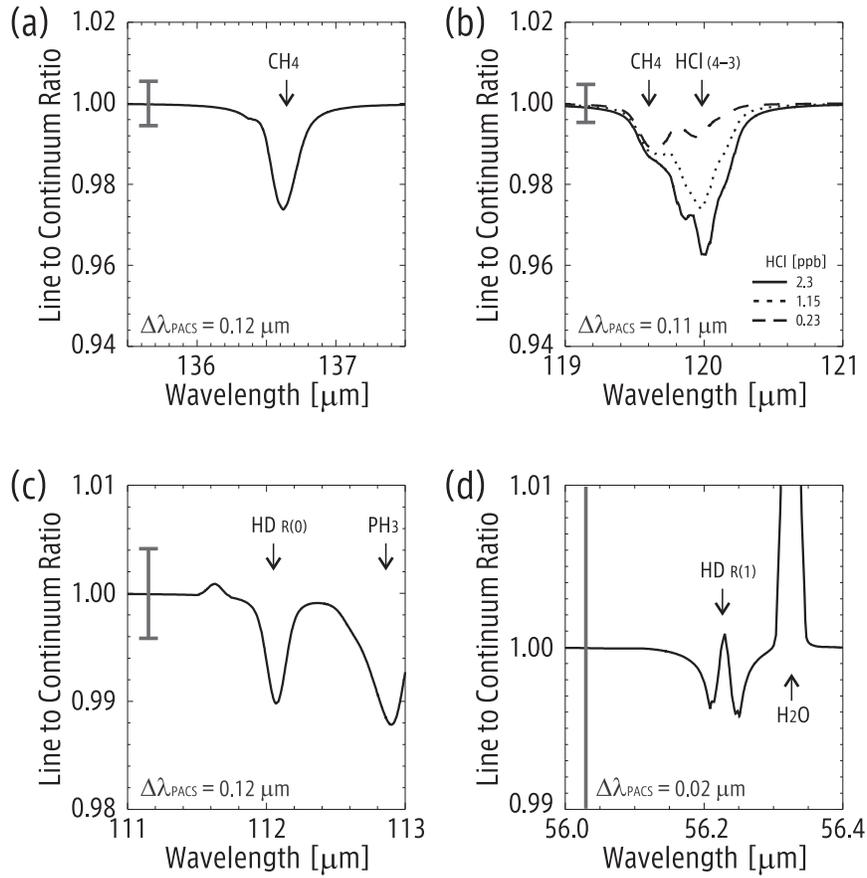}\\
\caption{Simulated PACS line-to-continuum ratio spectra for the minor species.
The gray bar at the left end of each spectrum indicates the 5-$\sigma$ levels of the assumed NESR.\@
Corresponding PACS spectral resolutions are indicated inside each panel. 
{\textbf (a)}: Close-up on the 136.6\,$\mu$m CH$_{4}$ line. 
{\textbf (b)}: Close-up on the HCl\,(4--3) transition.\@
The bold solid, dotted, dashed lines represent the cases of HCl abundance of 0.23, 1.15, and 2.3\,ppb, respectively. 
{\textbf (c)} \& {\textbf (d)}: Close-up on the HD rotational transitions. 
}
\label{minor}
\end{center}
\end{figure}

\clearpage

\begin{figure}
\begin{center}
\includegraphics{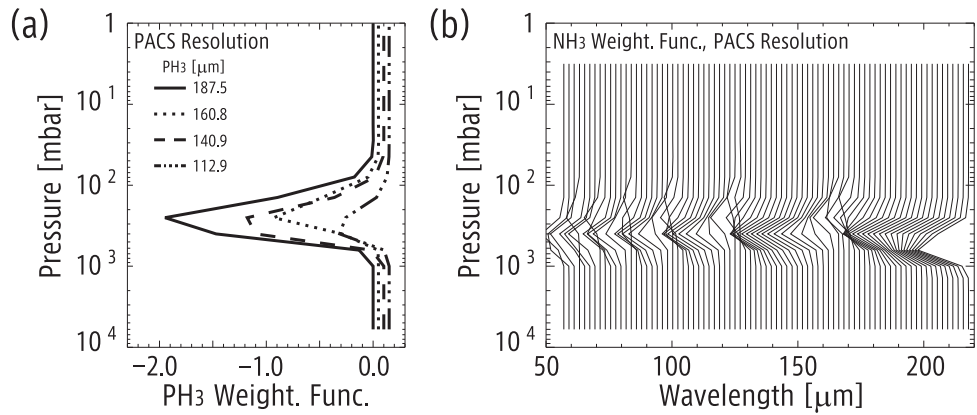}\\
\caption{Weighting functions with respect to {\textbf (a)} PH$_{3}$ and {\textbf (b)} NH$_{3}$ after considering the PACS spectral resolution.
For PH$_{3}$, 4 lines which are indicated by the arrows at the lowermost panel of Fig.~3 are considered.
For NH$_{3}$, the weighting functions along the PACS spectral range are plotted.}
\label{PH3_NH3_WF}
\end{center}
\end{figure}

\end{document}